\documentclass[english,10pt,twocolumn,final]{article} 

\usepackage{amsmath}
\usepackage{amssymb}
\usepackage{amsthm}
\usepackage{babel}
\usepackage{color}
\definecolor{darkblue}{cmyk}{0.83,0.89,0.0,0.43}
\usepackage{graphicx}
\usepackage[colorlinks,linkcolor=darkblue,urlcolor=blue,citecolor=darkblue,hyperindex,plainpages=false]{hyperref} 
\usepackage[T1]{fontenc}
\usepackage[latin1]{inputenc}
\usepackage{latex8}
\usepackage{times}

\makeatletter

\newcommand{\dcsFigRef}[1]{Fig.~\ref{fig:#1}}

\def\keywordname{{\bf Keywords:}}
\newcommand{\keywords}[1]{\par\addvspace\baselineskip
\noindent\keywordname\enspace\ignorespaces\textnormal{#1}}

\makeatother

\begin{document}

\bibliographystyle{latex8}

\title{The ShareGrid Portal: an easy way to submit jobs on computational Grids~\thanks{This work has been supported by TOP-IX and the Piedmont Region Agency under the Innovation Development Program.}}
\author{Cosimo Anglano \and Massimo Canonico \and Marco Guazzone \\
Department of Computer Science, University of Piemonte Orientale, Alessandria (Italy), \\
email:\{cosimo.anglano,massimo.canonico,marco.guazzone\}@unipmn.it
}

\maketitle

\thispagestyle{empty}


\begin{abstract}
Grid computing is a distributed computing paradigm which aims to aggregate several heterogeneous and distributed resources, belonging to different and independent organizations, in a dynamic, transparent and coordinated way.
Since its introduction, Grid computing has been successfully applied to solve several scientific challenging applications.
Despite of the consolidation of many of its aspects, there are some issues that are still open.
One of them is the transparency: in many real Grid systems, users still need to be aware of Grid computing, either for adapting their applications to this paradigm or for wrapping them in a suitable software framework.
In this paper we present the ShareGrid Portal, a Web portal and a portal framework, built on top of the ShareGrid project infrastructure.
Its intent is both to ease the execution of user applications in a Grid system and to allow developers to flexibly add new portal functionalities.
In this work, we compare it with other well-known Grid portals and we show its user interface and its architecture.
Finally we discuss user experiences and future extensions.


\end{abstract}



\section{The ShareGrid Portal} \label{sec:sgportal}


\subsection{Introduction} \label{ssec:intro}

With the advances of technology used for building scientific instrumentation, scientists are now able to explore aspects that only few years ago were not completely able to observe.
As a consequence, scientific applications, used to carry on scientific experiments, might need to access and elaborate massive amounts of data at a faster rate, and, therefore, they are characterized by an high demand of storage and computational resources.
For instance, the \emph{Large Hadron Collider} (LHC) \cite{LHC_URL}, the world's largest and highest-energy particle accelerator built by \emph{CERN} \cite{CERN_URL}, is able to produce petabytes of data per year that need to be accessed and studied by thousands of scientists belonging to different organizations and spread around the world.
The computing paradigm traditionally used for running compute-intensive applications is the \emph{cluster computing} \cite{Baker1999Cluster}, which implies the use of a powerful cluster of homogeneous computers owned by a single organization.
This paradigm does not fit very well with the above scenario, since the available resources might be insufficient to satisfy the incoming demand of computational power and storage space, and it generally does not allow collaboration among users of different organizations.
Hence, a more complex computing paradigm is needed in order to allow multi-institutional collaboration and resource sharing in a way as easy as possible.
The Grid computing \cite{Foster1998Grid} is a computing paradigm that tries to achieve this goal through a software and hardware infrastructure in order to allow and ease the dynamic sharing of heterogeneous resources among different, independent and geographically distributed organizations, in a way that should be totally transparent to the user.

The coordination and the sharing of heterogeneous resources between different communities distributed on a large geographic scale, has partially changed the way of solving compute-intensive and data-intensive applications.
Being a kind of a distributed system, one of the initial promise of the Grid computing was to make the execution of applications in a Grid system completely transparent from the point of view of users: the user of such a system should be unaware of the location where her/his application is currently executed; it should believe that all the computations be local.
This goal has been partially reached.
Even though a Grid middleware has the purpose of hiding all the low level interactions taking place in a Grid system, in practice the user has still the responsibility to describe the structure of her/his application or, in the worst case, adapts it to the requirements of the underlying Grid.
From the \emph{ShareGrid} \cite{Anglano2008CCGRID} project experience, we have learned that one of the most important aspect to consider for making a Grid project successful, is the social aspect, that is inducing a potential Grid user to use the Grid infrastructure as an everyday part of her/his work.

With this in mind, the ShareGrid project provides to its users community a Web portal, the \emph{ShareGrid Portal}, in order to ease the execution and the monitoring of a user application in the Grid system.

The ShareGrid Portal is a Web based Grid portal that provides to its users the access to Grid services and resources through the ShareGrid middleware.
Due to the nature of the current ShareGrid middleware, its primary focus is toward \emph{Bag-of-Tasks} (BoT) \cite{Cirne2003Running} applications.
A BoT application is a kind of parallel application whose tasks are independent from each other.
Though this is one of the simplest kind of programming model for the Grid computing, there are several real world applications that adopt it.
BoT applications include, but are not limited to, \emph{parameter sweep applications} \cite{Casanova2000Apples}, which are applications structured as a set of multiple experiments each of which is executed with a distinct set of parameters.
Parameter sweep applications can be viewed as a simple means of exploring the behaviour of a complex system through a series of parametric experiments.
Monte-Carlo and discrete-event simulation are a typical example of parameter sweep applications.
The wide applicability of this type of application contributed to the spread of the Grid computing in the scientific community \cite{Abramson2000IPDPS}.
In fact it is extensively used for carry on several experiments in many scientific areas such as: extra-terrestrial intelligence \cite{SETI_URL}, protein folding \cite{FAH_URL}, high-energy physics \cite{LCG_URL}, just to name a few.

The remainder of this section is organized as follows.
In \S\ref{ssec:related} we provide a short list of the most important related works.
In \S\ref{ssec:jsubmit} we explain how the job submission works, both with and without the portal, and what benefits the ShareGrid Portal brings.
In \S\ref{ssec:arch} we give an overview of the ShareGrid Portal architecture.
Finally, in \S\ref{ssec:arch} we present future works.



\subsection{Related Works} \label{ssec:related}

In this section we present some of similar and consolidated projects that are well-known to Grid community.
The \emph{GridSphere} project \cite{Novotny2004GridSphere} is a portal framework which provides a \emph{portlet}-based Web portal \cite{Abdelnur2003Portlet}; it supports various middlewares, like the \emph{Globus toolkit} \cite{Foster1997Globus}, \emph{Unicore} \cite{Romberg2000Unicore}, and \emph{gLite} \cite{Laure2004Glite}, through portlet components, called \emph{GridPortlets}.
The ShareGrid Portal is a portal framework too.
The main difference is the mechanism used for supporting a new Grid middleware.
While the GridSphere approach makes use of portlet components for extending the range of supported middlewares, the ShareGrid Portal extension mechanism consists in a series of \emph{Plain Old Java Object} (POJO) \cite{POJO_URL} interfaces, defining the high-level behaviour of a middleware, which are deployed in the portal by means of simple Java libraries (JARs).
Another difference is that GridSphere delegates each middleware portlet for providing its user interface, while the ShareGrid Portal provides a uniform view independent by the underlying middleware.

The \emph{P-GRADE Portal} \cite{Kacsuk2005Multi} is a workflow-oriented Grid portal; it is built upon GridSphere, for the Web interface, \emph{JavaGAT} \cite{Nieuwpoort2002Ibis}, for interacting with the middleware, and \emph{Condor DAGMan} \cite{DAGMAN_URL} for managing a workflow application.
The main difference with the ShareGrid Portal is the nature of the supported Grid applications; the P-GRADE Portal is oriented to workflow applications, with some extensions for parameter sweep applications; the ShareGrid Portal currently supports BoT applications, which include the family of parameter sweep applications but are more limited than workflow applications.
Another difference is the type of the supported middleware.
The P-GRADE Portal is a Globus-based, multi-Grid collaborative portal; it is able to connect to different Globus-based Grid systems and let their user communities to migrate applications between Grids.
The ShareGrid Portal can be considered a multi-Grid portal as well; however, it is not limited to the Globus middleware, since it relies on a set of interfaces that abstract from the underlying middleware; regarding the multi-Grid collaboration aspect, it is relied upon the underlying middleware.



\subsection{Job Submission} \label{ssec:jsubmit}

In this section we provide an overview of how the job submission works and the main motivations that brought us to develop a Web portal.

In order to submit an application to a Grid system, a user has to ``prepare'' a job that describes the structure and the behaviour of that application.
In \emph{OurGrid} version $3$ \cite{Cirne2003Running}, the middleware that the ShareGrid infrastructure currently adopts, the submission of a job is done through the \emph{mygrid} program \cite{Costa2004Mygrid}.
This is a Linux console application, installed on the user machine, that acts as a Grid local scheduler: it accepts in input a job file (i.e., the description of a user application) from the user and assigns it one or more computational resources according to a preconfigured scheduling policy.
The file representing the user job is a text file following the \emph{Job Description File} (JDF) format.
Basically, a JDF file is a text file containing a collection of task specifications, each of which includes an optional ``init'' section (for the stage-in phase), a ``remote'' section (for the remote execution phase) and a ``final'' section (for the stage-out phase), as shown in \dcsFigRef{sgportal-jdf}.
From the user point of view, the job submission phase requires three steps: (1) the user creates a job by writing a JDF file that provides the structure and the behaviour of the application (s)he wants to run, (2) the JDF file is submitted to the mygrid program, which transparently takes care of mapping the related job on one or more computational resources, and (3) the user manually and periodically polls the mygrid program for checking the job execution status.
\begin{figure}[htp]
\centering
\scriptsize{
\begin{verbatim}
job:
  label: MyJob
  requirements: mem == 100MB

...

task:
  init:
    put MyLocal.in MyRemote.in
    ...
  remote: MyCommand -i MyRemote.in -o MyRemote.out
  final:
    get MyRemote.out MyLocal.out
    ...
\end{verbatim}
}
\caption{ShareGrid Portal -- Example of a JDF file.}
\label{fig:sgportal-jdf}
\end{figure}
This workflow has some weak points:
\begin{itemize}
\item The mygrid program actually only runs on the Linux operating system; maybe, this is the major weakness since the computer world is not Linux-centric.
\item The mygrid program maintains its state in the volatile memory of the client machine and so it must remain running for the entire duration of the job execution.
For this reason, the user must keep her/his machine powered on until all of her/his jobs in execution are done.
This point represents another important weakness causing both economical and ecological implications for the useless power consumption.
In fact, the price of the energy continuously gets higher and higher due to the strong dependence to oil and to the current oil crisis caused by an unbalance between the demand and the offer.
Hence, a possible economical consequence for a user is the increasing of her/his energy costs.
For what concerning ecological implications, it is now a fact that the excessive energy consumption contributes to the global warming.
Users that are sensible to this subject might be disappointed for seeing that, in some sense, they are contributing to the overheating of the planet Earth.
\item The mygrid program is not free in resource occupation.
We have empirically observed that when the mygrid program is in an idle state, the memory consumption is about of $30$MB (the CPU, on the other hand, is nearly unutilized).
For this reason, the user might consider the installation and the use of the mygrid program as something of too intrusive, especially when (s)he does not own a powerful machine.
\item The user must repeatedly poll the mygrid application for knowing the execution status of a given job (e.g., either completed, failed or still running).
\item The JDF syntax, though simple, is error-prone, especially for the beginner user; furthermore, some kind of error (e.g. a misspelled file name) might only be thrown near the end of the execution, making the entire computation useless.
\end{itemize}
The above issues were enough to motivate the realization of the ShareGrid Portal.
The first benefit the Web portal brings is the operating system independence, freeing the user from having installed on her/his own machine the Linux operating system.
A Web application has also the advantage that does not require any software installation, since almost all modern operating system distributions ship with a Web browser and the utilization of the resources of the user machine is rather limited.
Because the Web portal does not store any job submission state on the user machine, the user can submit her/his jobs from any machine connected to the Internet; furthermore, there is no more need to keep the machine powered on, waiting for jobs completion, since all the informations about job submissions are kept in the ShareGrid infrastructure.
In order to avoid the user to manually and periodically poll for monitoring the status of the job execution, the ShareGrid Portal provides an active notification system.
When the execution status of a job changes (e.g. from running to finished), the portal sends a notification (actually an email) to the user.

\begin{figure}[htp]
\centering
\includegraphics[scale=0.75]{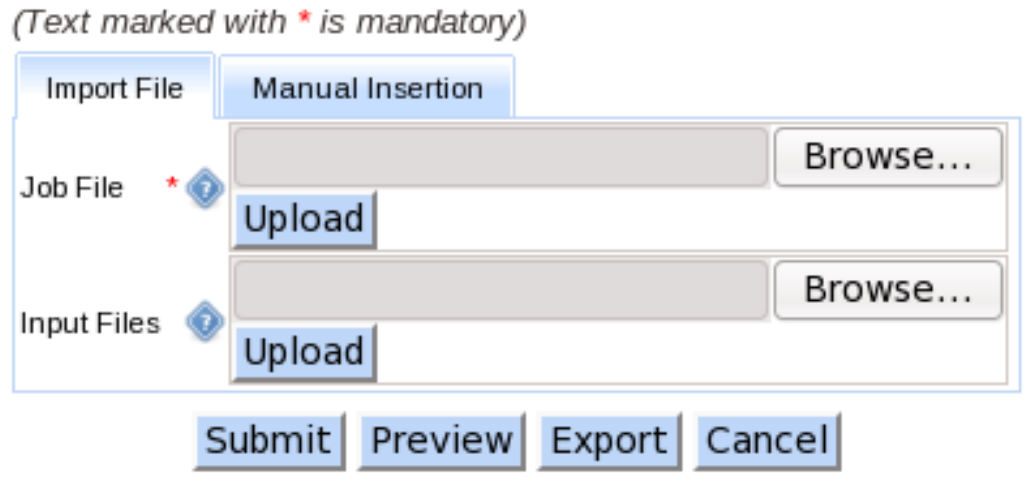}
\caption{ShareGrid Portal -- Job submission through job file import.}
\label{fig:sgportal-jobimpjdf}
\end{figure}

\begin{figure}[htp]
\centering
\includegraphics[scale=0.6]{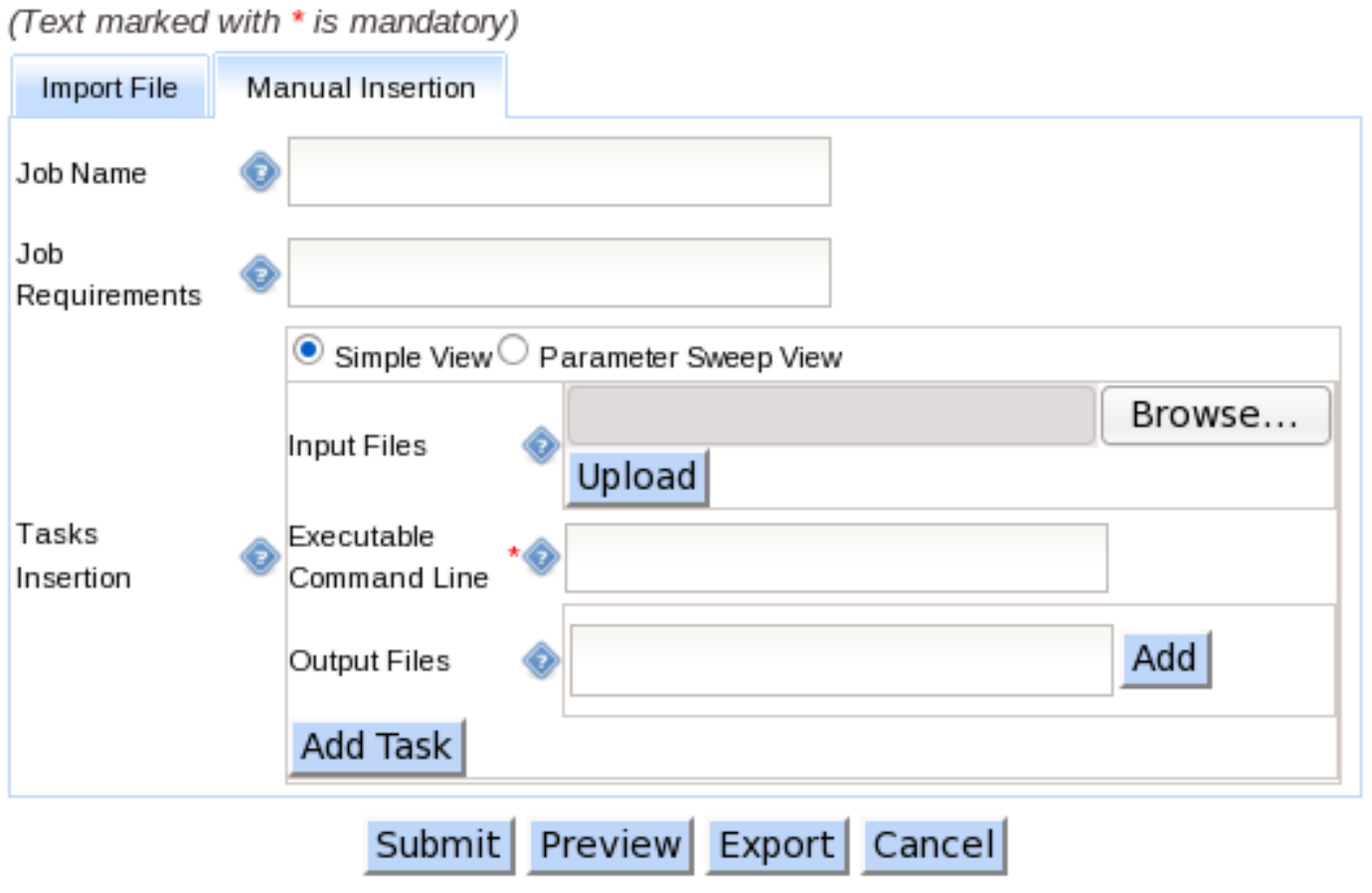}
\caption{ShareGrid Portal -- Job submission through the simple view.}
\label{fig:sgportal-jobsimple}
\end{figure}

\begin{figure}[htp]
\centering
\includegraphics[scale=0.6]{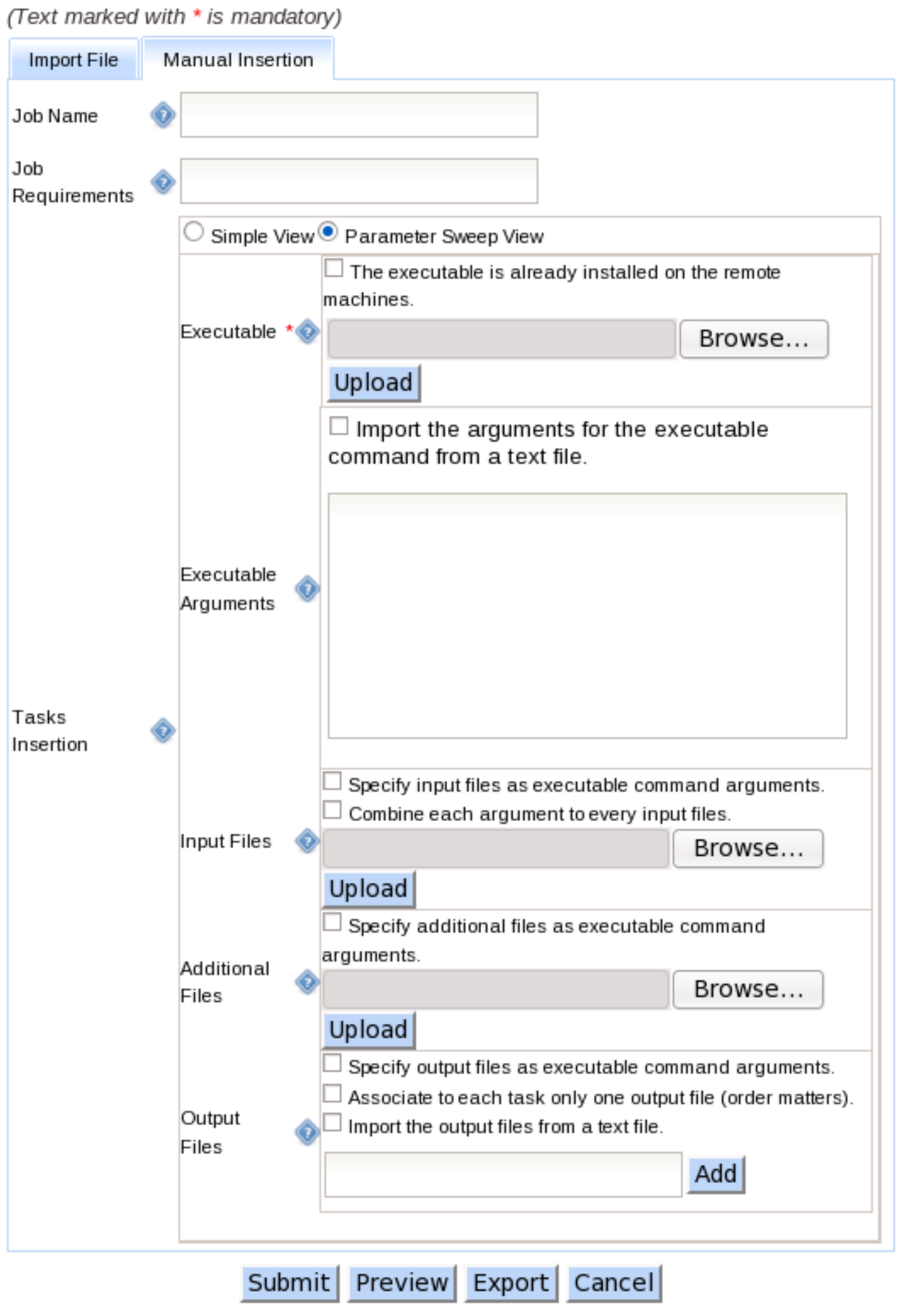}
\caption{ShareGrid Portal -- Job submission through the parameter sweep view.}
\label{fig:sgportal-jobpsa}
\end{figure}

The submission of a job has been simplified thanks to the presence of several user-friendly web interfaces.
These interfaces divides in two main groups according to the way job informations are fed: (1) job file upload and (2) manually job insertion.
The fastest way to submit a job through the portal is by using the ``import file'' interface, depicted in \dcsFigRef{sgportal-jobimpjdf}.
This interface allows the user to directly upload a JDF job file, that is a file that contains informations about the structure and the behaviour of the user application.
Along with the job file, the user can upload many input files as needed that will be transfered on worker machines during the stage-in phase.
This type of job submission is targeted to expert users that do not want to go through the additional steps that are inevitably introduced by the others more user-friendly interfaces.
The other type of job submission interface is the manual job insertion view.
In this interface, the user can choose to insert her/his job between two views: (1) a generic simple interface and (2) an ad-hoc interface for parameter sweep applications.
In the generic simple interface, shown in \dcsFigRef{sgportal-jobsimple}, the user can create a job for a generic BoT application.
For each task, the user must explicitly specify the executable command line (i.e., the executable name and its argument that will be executed on the worker machine), can optionally upload input files (included the executable command if not already present in the worker machines) and possibly specify one or more output file names.
This last two informations are used during the job stage-in and stage-out phase, respectively.
The other way for manually submit a job is the parameter sweep view, shown in \dcsFigRef{sgportal-jobpsa}, an ad-hoc interface targeted for parameter study applications.
This kind of application differs from a generic BoT application for the executable command: possibly different for each task in a BoT application and unique in a parameter sweep application.
In fact, in this interface, the user is asked to specify a unique executable command, the list of parameters to study (one line for each experiment) and zero or more input, output and shared files.
In order to speed up the insertion of informations, this view provides many useful shortcuts; the ones that are worth noting are:
\begin{itemize}
\item The user can choose to manually insert the executable command name, in the case it is already installed on worker machines, or to upload the corresponding executable command file.
\item The arguments for the executable command and the output file names can be either uploaded via a text file or inserted by hand.
\item Parameters can be studied as a function of input files by combining each parameter line to every input files.
\end{itemize}
Each of the above interfaces allows the user to preview the job before submitting it, for discovering possible syntax errors, and to export the job to a text file (actually, only to a JDF file); the exported job can be successively uploaded in the import file view, in order to let the user to minimize repetitive tasks for similar jobs.
After a job has been submitted, the user is freed from any other task and can decide, for example, to submit another job or even to shutdown her/his machine.
It is the responsibility of the ShareGrid Portal to instruct the underlying Grid middleware for staging in the input files, remotely executing the specified command and, finally, for staging out the output files.
In particular, staged out files are stored in the ShareGrid repository, that is an area accessible only to the user who submitted the job.
Once the execution status of a job is changed (e.g. from running to finished or to failed), the ShareGrid Portal sends a notification to the user.
For instance, when a job has been successfully completed, the notification includes a link to the ShareGrid repository where the output files has been stored.



\subsection{Architecture} \label{ssec:arch}

In this section we provide a high-level description of the ShareGrid Portal architecture.
The ShareGrid Portal is both a Web portal application and a portal framework.

As a Web portal application, it provides internationalization and localization support, user account management, data persistence abstraction, graphical appearance customization, and a set of core functionalities for the creation, deletion, updating and querying of user and job informations.
The access control to the portal is based upon the \emph{Role Based Access Control} (RBAC) model \cite{Ferraiolo1992Role}.
As pointed out in \cite{Sandhu1996Role}, a role is a semantic construct around which access control policies are formulated; users are assigned to specific roles and, in this way, they acquire the permissions associated with their roles.
Roles are closely related to the concept of groups but the main difference is that a role brings together a set of users on one side and a set of groups on the other, while a group is typically defined as a set of users.
Our RBAC model consists of a hierarchical user role model where each role is assigned to one or more access permissions defined upon a hierarchical page access control policy.
For instance, the administrator user is allowed to access to any page whereas the anonymous user can access only to a limited set a pages (e.g., to the user registration page).
Actually, we have defined three roles: (1) Anonymous, for users not logged into the portal, (2) Standard, for users that are allowed to submit a job to the ShareGrid infrastructure, and (3) Administrator, for standard users with additional site administration privileges.

\begin{figure}[htp]
\centering
\includegraphics[scale=0.3]{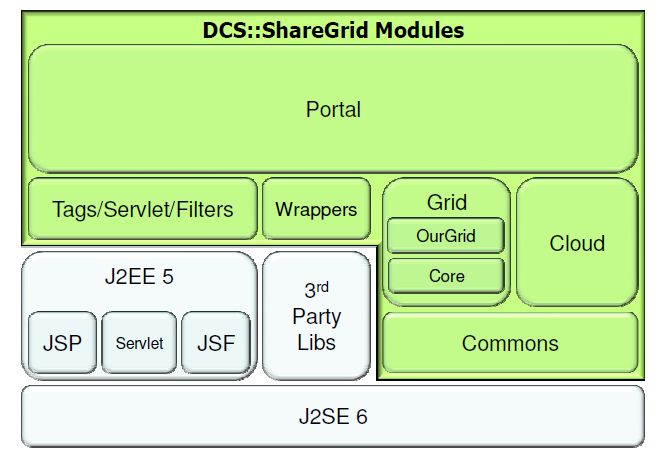}
\caption{ShareGrid Portal -- High level view of the architecture.}
\label{fig:sgportal-archblock}
\end{figure}

As a portal framework, the ShareGrid Portal provides to the developer a set of independent and reusable components.
In \dcsFigRef{sgportal-archblock} is depicted an high-level view of the ShareGrid Portal infrastructure.
All the modules relies on the \emph{Sun Java Platform Standard Edition $6$} \cite{J2SE_URL}.
The ``Commons'' module offers shared and commonly used functionalities, like string manipulation, format conversions, I/O and network utilities, and so on.
This component is used by almost any other ShareGrid modules.
The ``Grid'' module aims to provide an abstraction layer from any Grid middleware; the ShareGrid portal uses this component for keeping it independent by the underlying middleware used.
This module is divided into two parts: the ``Core'' sub-module defines the interfaces and the implementations that are middleware independent, whereas the OurGrid sub-module is the implementation for the OurGrid middleware.
The role of the ``Portal'' module is two fold: (1) it provides a set of interfaces, classes and tag libraries to act as a Web application framework for developing \emph{Sun Java EE} \cite{J2EE_URL} Web applications, and (2) it realizes the Web interfaces for using it as a Grid portal.
A Grid portal derived from the ``Portal'' module, included the ShareGrid Portal, consists of at least a set of presentation pages, including static \emph{HTML} \cite{Raggett1999Html401}, \emph{Sun JavaServer Pages} \cite{JSP_URL} and \emph{JavaServer Faces} \cite{JSF_URL}, along with the associated backing beans, for implementing the presentation logic; in addition, it is possible to define classes for realizing the business logic and overriding existing classes for redefining, for instance, the data persistence layer or the page life cycle.



\subsection{Conclusions and Future Work} \label{ssec:conclusions}

The ShareGrid Portal is a rather young project started in the middle of 2007.
Nevertheless, it is able to provide a simple but effective way to submit jobs to the ShareGrid infrastructure avoiding to force its users to adapt their desktop environment to the requirements of the underlying middleware.
Obviously, from the point of view of an expert user, the time taken for submitting a job to the Grid middleware with the portal will never be comparable with the one spent directly using the console application.
In fact, this is a trade-off that almost all Web applications have to accept with respect to the desktop-based counterparts.
However, we think the benefits brought by a Web portal might make the adoption of the ShareGrid infrastructure more attractive.

Being a young project it is in continue evolution.
Ongoing projects include the redesign of some views, in order to make the job submission even faster, and the development of application oriented Web interfaces, that is interfaces specifically targeted to an application domain, like distributed rendering.
Future extensions include the support for others Grid middlewares, the implementation of a Web Services layer and the possibility to export a job to different formats, like the Job Submission Description Language (JSDL) \cite{Anjomshoaa2008JSDL} format.



{\small
\bibliography{paper}
}


\end{document}